\documentclass[nofootinbib,reprint,superscriptaddress,preprintnumbers]{revtex4-2}

\usepackage[english]{babel}

\usepackage[letterpaper,top=2cm,bottom=2cm,left=3cm,right=3cm,marginparwidth=1.75cm]{geometry}

\usepackage{amsmath}
\usepackage{graphicx}
\usepackage{url}
\usepackage[colorlinks=true, allcolors=blue]{hyperref}
\usepackage{enumerate}
\usepackage{bm}
\usepackage{comment}

\usepackage[condensed]{roboto}

\begin{document}

\preprint{LA-UR-25-25776}

\title{Qubit thermodynamics: Entropy production from nonadiabatic driving}

\author{Pavel Zhelnin}
\affiliation{Department of Physics \& Laboratory for Particle Physics and Cosmology, Harvard University, Cambridge, MA 02138, USA}

\author{Lucas Johns}
\email{ljohns@lanl.gov}
\affiliation{Theoretical Division, Los Alamos National Laboratory, Los Alamos, NM 87545, USA}

\author{Carlos A. Argüelles}
\affiliation{Department of Physics \& Laboratory for Particle Physics and Cosmology, Harvard University, Cambridge, MA 02138, USA}


\begin{abstract}
Adiabaticity is a cornerstone of many promising approaches to quantum control, computing, and simulation.
In practice, however, there is always a trade-off. 
Although the deleterious effects of noise can be diminished by running a control schedule more quickly, this benefit comes at the expense of nonadiabaticity. 
To put these two unwanted effects on the same theoretical footing, we analyze the nonadiabatic error in qubit control as a form of entropy production, examining the mechanism by which fine-grained information is effectively lost despite the dynamics being fundamentally unitary.
A crucial issue here is the question of how to define equilibrium under a time-dependent Hamiltonian.
Using the Landau--Zener protocol as a test case, we show that entropy increases nearly monotonically when equilibrium is defined with respect to the effective Hamiltonian in the optimal superadiabatic frame.
We then consider single-passage Landau--Zener--St\"{u}ckelberg--Majorana interferometry, in which the initial state of the qubit is arbitrary.
Violations of the second law of thermodynamics are possible but require exquisite control to achieve deliberately.
\end{abstract}

\maketitle

\section{Introduction}

Adiabaticity is foundational to many techniques currently being developed for the control of quantum systems. 
Prominent examples include adiabatic quantum computing~\cite{albash2018}, (accelerated) adiabatic gates~\cite{wu2013, martinis2014, huang2019, ribeiro2019, setiawan2021}, adiabatic pulses in magnetic resonance~\cite{deschamps2008, vandermause2016}, adiabatic operation of quantum heat engines~\cite{bender2000}, and the framework of superadiabaticity, transitionless driving, and shortcuts to adiabaticity~\cite{berry1987, berry1990, lim1991, berry2009, bason2012, guery2019}.

In adiabatic control protocols, the overarching idea is to take advantage of the robustness of adiabaticity.
Strictly speaking, however, the adiabatic theorem requires an infinitely slow process.
In practice, this exposes a quantum system to noise and environment-induced fidelity loss.
On the other hand, a more rapid control protocol induces nonadiabaticity by inciting transitions between the energy levels of the system.
Nonadiabaticity equates to erroneous control even if the evolution is perfectly unitary because the system does not end up in the desired final state, \textit{i.e.}, the one that would be achieved if the evolution were fully adiabatic.

In the open-system dynamics of a qubit, decoherence results in entropy production.
Fundamentally, information is lost to the environment.
In this work, we are motivated by the development of a unified thermodynamic treatment of the tradeoff between environmental decoherence and nonadiabaticity.
To this end, we focus on entropy production caused by nonadiabatic driving.
Information loss in this case is associated with a fine-grained--coarse-grained split rather than a system--environment distinction.
The mechanism of entropy increase is dephasing, with the phase information in question being effectively lost because it is unresolved or practically inaccessible.
Within this approach, the quantum-mechanical adiabaticity of a driven, isolated system is equivalent to thermodynamic adiabaticity. 
At a finite rate of driving, the coarse-grained system ``heats up'' as the actual fine-grained state becomes less certain.

In quantum thermodynamics, entropy production from nonadiabatic driving falls under the rubric of irreversible work~\cite{campisi2011, landi2021}.
Studies typically focus on the entropy change $\Delta S \equiv S(t_f) - S(t_i)$ as a system is driven over a time period $\Delta t \equiv t_f - t_i$, with the system rethermalized at the end. 
In this work we formulate thermodynamic notions---equilibrium, entropy, and so on---in a manner suited to an ongoing control protocol, inspired by the continuous production of entropy from an ongoing interaction between system and environment in open-system evolution (\textit{e.g.}, Lindblad dynamics). 
To what extent is entropy monotonically produced by nonadiabatic driving, with fine-grained information continuously lost?

In attempting to address this question, we encounter the issue that there are no unique thermal equilibrium states under a time-dependent Hamiltonian $H(t)$ since the energy eigenstates are changing with time.
Entropy inherits this ambiguity. 
The path from $S(t_i)$ to $S(t_f)$ depends on which time-dependent frame is adopted.

In the following, we define \textit{frame-dependent} equilibrium ensembles of coarse-grained unitary evolution.
Then, focusing on the analysis of the Landau--Zener protocol~\cite{shevchenko2010, silveri2017, ivakhnenko2023}, we show that nonadiabatic driving exhibits highly non-monotonic entropy production when the coarse-grained entropy is formulated in the ordinary adiabatic basis of instantaneous energy eigenstates of the lab-frame Hamiltonian.
In the optimal superadiabatic frame~\cite{berry1987, berry1990, lim1991}, however, the accumulation of entropy is nearly monotonic.
This is our central idea: the second law of thermodynamics $dS/dt \geq 0$ is not strictly satisfied, but it is much more nearly obeyed in some frames than others.
In highlighting superadiabatic frames, we are building in particular on Refs.~\cite{deschamps2008, vandermause2016}, which found superadiabaticity to be useful in shaping quantum-control pulses.

This study is motivated by the technical challenge of addressing qubits, but it is worth noting that close parallels arise in some rather disparate areas of physics. 
For example, particle production in an expanding universe or time-dependent electric field faces the issue that particle number is ambiguously defined (\textit{i.e.}, it is frame-dependent).
The virtues of superadiabatic frames have previously been explored in these contexts~\cite{dabrowski2014, dabrowski2016, yamada2021, alvarez2022}.
Another setting where related issues arise is neutrino transport in stellar environments, where neutrinos propagate through a slowly varying astrophysical medium~\cite{raffelt1996stars, giunti2007fundamentals, johns2025neutrino, 
volpe2024neutrinos}.
The quantum thermodynamics of coherent neutrino flavor evolution is just beginning to be explored~\cite{johns2023thermodynamics, fiorillo2025dispersion, johns2025local}.
The Landau--Zener protocol appears in this field as well~\cite{parke1986, haxton1986}, with quantum control effectuated by the astrophysical medium.

In Sec.~\ref{sec:thermo} we introduce the coarse-grained entropy and other relevant notions pertaining to quantum thermodynamics and quantum information theory.
In Sec.~\ref{sec:LZ} we analyze entropy production in Landau--Zener driving with an emphasis on the frame-dependence.
Then, in Sec.~\ref{sec:arbitrary} we consider driving with arbitrary initial states, as in Landau--Zener--St\"{u}ckelberg--Majorana interferometry~\cite{ivakhnenko2023}. 
We show that, while entropy reduction is achievable, the change in entropy is highly sensitive to the control parameters.
In Sec.~\ref{sec:discussion} we conclude.

\section{Qubit thermodynamics\label{sec:thermo}}

To begin, we review some relevant concepts from quantum thermodynamics and quantum information theory and discuss entropy growth under projection and coarse-graining.
Through the rest of this article, we set $\hbar = k_B = 1$.

We will be making extensive use of the following decomposition of $\rho$ for a two-level system:
\begin{equation}
    \rho = \frac{P_0+\bm{P}\cdot\bm{\sigma}}{2}. \label{eq:pauli}
\end{equation}
Here, $P_0 = 1$ is the trace of $\rho$, $\bm{P}$ is the qubit Bloch vector, and $\bm{\sigma} = ( \sigma_x, \sigma_y, \sigma_z)$ is a vector of Pauli matrices.
We also introduce $H_0$ and $\bm{H}$ through an identical decomposition for the Hamiltonian. Using Eq.~\eqref{eq:pauli}, the von Neumann equation implies the equation of motion
\begin{equation}
\frac{d\bm{P}}{dt} = \bm{H} \times \bm{P}, \label{eq:precession}
\end{equation}
which entails precessional motion of $\bm{P}$ around $\bm{H}$, similar to a magnetic moment precessing in a magnetic field.

Suppose the system is weakly coupled to an environment at temperature $T = \beta^{-1}$.
The Gibbs equilibrium is
\begin{equation}
    \rho^\textrm{eq} = \frac{e^{-\beta H}}{Z}, \label{eq:gibbs}
\end{equation}
where $H$ is the system Hamiltonian and $Z$ is the partition function, $Z = \textrm{Tr} ( e^{-\beta H} )$.
The state $\rho^\textrm{eq}$ maximizes the von Neumann entropy $S = - \textrm{Tr} ( \rho \log \rho )$ at fixed expectation value of internal energy, $E = \textrm{Tr} ( H \rho )$.

The equilibrium Bloch vector, defined from $\rho^{\textrm{eq}}$ via Eq.~\eqref{eq:pauli}, is
\begin{equation}
    \bm{P}^\textrm{eq} = - \tanh \left( \frac{\beta |\bm{H}|}{2} \right) \bm{\hat{H}}. \label{eq:equilP}
\end{equation}
For $T > 0$, $\bm{P}^\textrm{eq}$ points in the opposite direction of $\bm{H}$, indicating greater occupation of the lower of the two levels.
Population inversion occurs at negative temperatures.
$\rho^\textrm{eq}$ converges on an equal mixture in the $T \rightarrow \infty$ limit with finite $|\bm{H}|$, implying full depolarization $\bm{P}^{\textrm{eq}} \cong 0$. 
Equilibrium approaches a pure state in the $T \rightarrow 0$ or $|\bm{H}| \rightarrow \infty$ limit. 
Just as the coherent motion [Eq.~\eqref{eq:precession}] resembles Larmor precession, the equilibrium polarization [Eq.~\eqref{eq:equilP}] resembles the equilibrium magnetization of the mean-field Ising model~\cite{callen1980thermodynamics}.

After plugging in Eq.~\eqref{eq:pauli}, the von Neumann entropy becomes
\begin{align}
    S(\bm{P}) = &- \left( \frac{1 + |\bm{P}|}{2} \right) \log \left( \frac{1 + |\bm{P}|}{2} \right) \notag \\
    &- \left( \frac{1 - |\bm{P}|}{2} \right) \log \left( \frac{1 - |\bm{P}|}{2} \right), \label{eq:entropy}
\end{align}
which is a decreasing function of $|\bm{P}|$.
Let $\langle \cdot \rangle_t$ denote the time average over some coarse-graining window.
Then $|\langle \bm{P} \rangle_t| \leq \langle |\bm{P}| \rangle_t$ and so, under Eq.~\eqref{eq:precession},
\begin{equation}
    S(\langle \bm{P} \rangle_t) \geq \langle S(\bm{P}) \rangle_t.
\end{equation}
Although we write the time-averaged entropy on the right-hand side, $S(\bm{P})$ is in fact constant over the averaging window---and throughout the entirety of the evolution---because we are assuming unitary dynamics. 
The coarse-grained entropy $S(\langle\bm{P}\rangle_t)$, on the other hand, may increase, decrease, or remain constant depending on the particular dynamics within the time-averaging window.

To clarify the question we raise in this study, it is helpful to draw a contrast with entropy production in an open quantum system.
Under certain assumptions about the system--environment coupling and the environment itself, the dynamics is such that the entropy is nondecreasing.
For example, suppose that $\rho$ evolves under a dynamical semigroup:
\begin{equation}
    \rho (t) = e^{\mathcal{L}t} \rho(0).
\end{equation}
The entropy production rate is
\begin{equation}
    \sigma (t) \equiv - \frac{d}{dt}S(\rho(t) || \rho(0)),
\end{equation}
where the quantum relative entropy is
\begin{equation}
    S(\rho || \rho') \equiv \textrm{Tr} (\rho \log \rho) - \textrm{Tr} ( \rho \log \rho').
\end{equation}
Under the dynamical semigroup, the entropy production rate is nonnegative at all times~\cite{breuer2002}:
\begin{equation}
    \sigma \geq 0. \label{eq:sigma0}
\end{equation}
To what degree does coarse-grained entropy production from nonadiabatic driving display similar behavior, with information continuously lost?

In contemplating this question, there are two competing intuitions.
On the one hand, if $\bm{\hat{H}}$ changes slowly relative to the precession frequency $|\bm{H}|$, then the correlation between the motion of $\bm{\hat{H}}$ and the position of $\bm{P}$ is relatively weak.
In the adiabatic limit, $\bm{P}$ precesses many times around $\bm{\hat{H}}$ as the latter moves.
Compare this scenario to the sudden limit: the change in $\bm{\hat{H}}\cdot\bm{P}$ over some period is highly dependent on the initial orientation of $\bm{P}$ with respect to $\bm{\hat{H}}$.
In other words, the adiabatic limit coincides with an insensitivity to the phase, at any given moment, of $\bm{P}$ with respect to $\bm{\hat{H}}$. 
This intuition hints at a sort of scrambling of phase information. 
On the other hand, the evolution under Eq.~\eqref{eq:precession} is unitary, and so there is no actual mechanism of irreversibility or information loss.

The instantaneous motion of $\bm{P}$ is to follow a circular orbit around $\bm{H}$. 
Thus $\bm{H}(t)$ defines an ensemble of qubit states at every moment $t$ consisting of all values of $\bm{P}$ visited along the circular orbit.
We call this set of states the equilibrium ensemble and denote it with the symbol $f^\textrm{eq}$.
If $\bm{H}$ changes infinitely slowly, then over any finite period
\begin{equation}
    \langle \bm{P} \rangle_t \cong \langle \bm{P} \rangle_\textrm{eq}. \label{eq:ergodic}
\end{equation}
Note the resemblance to the property of ergodicity, with equality between the time average and $f^\textrm{eq}$ ensemble average.
Using the right-hand side, we define the instantaneous equilibrium state under coherent evolution
\begin{equation}
    \bm{P}^\textrm{eq} \equiv \langle \bm{P} \rangle_\textrm{eq} = s|\bm{\hat{H}}\cdot\bm{P}|\bm{\hat{H}}, \label{eq:Peq}
\end{equation}
where $s = \pm 1$ is the alignment factor. 
The more adiabatic the driving is, the better the approximation in Eq.~\eqref{eq:ergodic} is.
This suggests that nonadiabatic entropy production accumulates due to $\bm{P}$ being continuously driven away from equilibrium, with equilibrium defined by Eq.~\eqref{eq:Peq} and equilibration constantly occurring due to Eq.~\eqref{eq:ergodic}.

The entropy produced in this situation is effectively a form of dissipated work~\cite{jarzynski1997, crooks1998, strasberg2022}.
If, after out-of-equilibrium driving, the system is in state $\rho^f$ and is then rethermalized to $\rho^\textrm{eq}$, the nonequilibrium entropy production is $S (\rho^f || \rho^\textrm{eq})$~\cite{kawai2007, vaikuntanathan2009, deffner2010}.
However, in our setup, there is no external environment to impose thermalization.
Entropy production occurs through effective equilibration via temporal coarse-graining: $\rho^{\textrm{eq}}$ is defined using Eqs.~\eqref{eq:pauli} and \eqref{eq:Peq} instead of Eq.~\eqref{eq:gibbs}.

In the following sections we study
\begin{equation}
    \Delta S (t) \equiv S(\bm{P}^{\textrm{eq}}(t)) - S(\bm{P}^{\textrm{eq}}(t_i)),
\end{equation}
where $t_i$ is the initial time. 
Growth of $\Delta S$ is associated with a decrease in $|\bm{P}^{\textrm{eq}}|$ and therefore with loss of purity:
\begin{equation}
    \mathcal{P} \equiv \textrm{Tr} ((\rho^{\textrm{eq}})^2) = \frac{P_0^2 + |\bm{P}^{\textrm{eq}}|^2}{2}.
\end{equation}
Greater $\Delta S$ over the duration of a control protocol also equates to worse fidelity 
\begin{equation}
    \mathcal{F} (\rho, \rho')  = \left( \textrm{Tr} \sqrt{\sqrt{\rho} \rho' \sqrt{\rho}} \right)^2.
\end{equation}
To restate our guiding question: To what degree does $\Delta S(t)$ monotonically increase under nearly adiabatic driving, paralleling Eq.~\eqref{eq:sigma0} for system--environment coupling?
Monotonic increase indicates that the performance of qubit control consistently deteriorates throughout the protocol.

\section{Landau--Zener entropy production\label{sec:LZ}}

From this point onward we develop our ideas in relation to a concrete example: the Landau--Zener protocol, the archetypal example of (nearly) adiabatic driving of a quantum system~\cite{shevchenko2010, silveri2017, ivakhnenko2023}.
The control protocol is defined by the Hamiltonian
\begin{equation}
    H(t) = \sigma_x + \epsilon t \sigma_z,
\end{equation}
where $\epsilon > 0$ is a constant. 
Note that we have adopted units in which $H_{12} = 1$. 
Henceforth we report quantities as dimensionless, with their dimensional values related to this choice.

The Hamiltonian vector is
\begin{equation}
\bm{H} (t) = 2\bm{\hat{x}} + 2\epsilon t \bm{\hat{z}}.
\end{equation}
As time progresses, $\bm{\hat{H}}$ sweeps over a curve beginning near the south pole of the Bloch sphere and ending near the north pole.
The qubit energy levels are
\begin{equation}
    E_{\pm}(t) = \pm |\bm{H}| = \pm 2 \sqrt{1 + ( \epsilon t)^2}.
\end{equation}
A resonance occurs at $t = 0$, which is also the time at which $\bm{\hat{H}}$ changes most rapidly.

For a qubit beginning in an energy eigenstate, the probability of a nonadiabatic transition occurring from initial time  $-\infty$ to final time $+ \infty$ is given by the Landau--Zener formula
\begin{equation}
    p_{\textrm{LZ}} = e^{-\pi / \epsilon}.
\end{equation}
This probability implies a certain asymptotic value of $|\langle \bm{P} \rangle|$ and therefore of $S$.
We use the symbol $\Delta S_{\textrm{LZ}}$ to label the entropy production implied by the Landau--Zener formula. 
It is calculated by plugging $|\bm{P}| = 1 - 2 p_{\textrm{LZ}}$ into Eq.~\eqref{eq:entropy}.

\begin{figure}
    \centering
    \includegraphics[width=\linewidth]{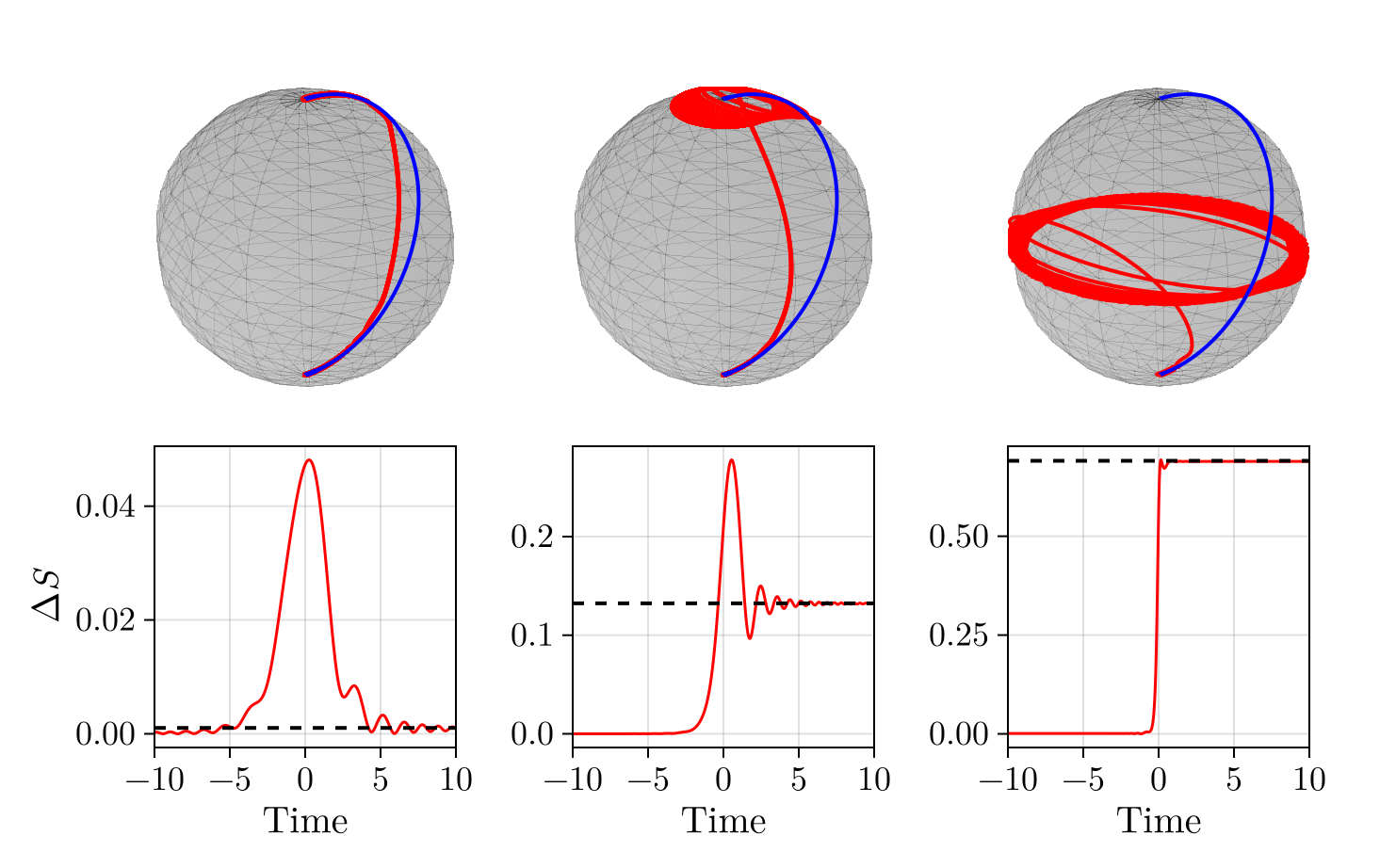}
    \caption{\textbf{\textit{Numerical solutions of Landau--Zener protocols in the lab frame.}}
    Results are shown for three choices of sweep-rate parameter $\epsilon$: $0.34$, $0.89$, and $5$ (from left to right). In the upper panels, red and blue curves show the Bloch-sphere trajectories of the Hamiltonian and qubit Bloch vectors $\bm{H}$ and $\bm{P}$. In the lower panels, the red curves show the coarse-grained entropy production $\Delta S$ over time and the dashed lines show the Landau--Zener asymptotic entropy $\Delta S_{\textrm{LZ}}$. Entropy production in the adiabatic control regime ($\epsilon \lesssim 1$) is highly non-monotonic.
    }
    \label{fig:bloch1}
\end{figure}

\begin{figure}
    \centering
    \includegraphics[width=\linewidth]{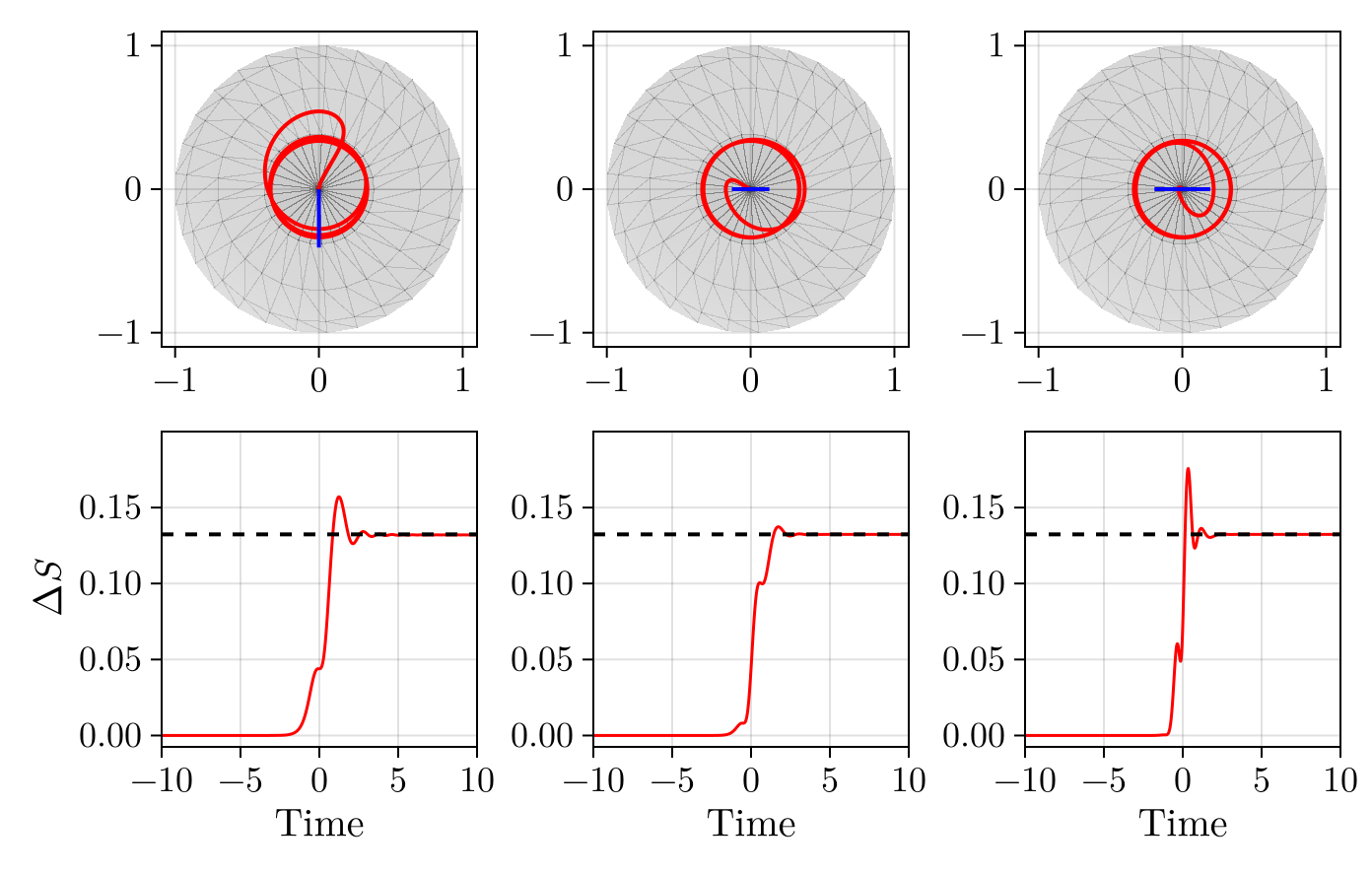}
    \caption{\textbf{\textit{Numerical solution of a Landau--Zener protocol in different superadiabatic frames.}} Trajectories and entropies are calculated in (from left to right) the 1st, 2nd, and 4th superadiabatic frames for the same control schedule as in the middle panels of Fig.~\ref{fig:bloch1}.
    The 1st and 4th frames show deviations from a nearly monotonic increase in $\Delta S$. 
    The 2nd frame is optimal and shows the most nearly monotonic $\Delta S$.
    }
    \label{fig:bloch2}
\end{figure}

In Fig.~\ref{fig:bloch1} we present numerical solutions of the Landau--Zener protocol for three representative choices of the parameter $\epsilon$.
Trajectories of $\bm{P}$ on the Bloch sphere are shown in the upper panels.
The smaller $\epsilon$ is, the better $\bm{P}$ is able to follow $\bm{H}$ as the latter sweeps from the south to the north pole.
The lower panels show the evolution of $\Delta S$.
In all cases $\Delta S$ increases from the beginning to the end of the protocol and asymptotically agrees with $\Delta S_{\textrm{LZ}}$.
In the adiabatic limit, however, the increase occurs in a highly non-monotonic fashion: after peaking at the resonance, $\Delta S$ drops sharply as it undergoes decaying oscillations around the asymptotic value. 
Based on this result, the conjecture that entropy is continuously produced---and never reduced---during the driving appears to be egregiously violated.

On the contrary, what we will now demonstrate is that $\Delta S$ does in fact become nearly monotonic upon switching to the optimal superadiabatic frame.
The machinery of superadiabatic frames originates with work by Berry~\cite{berry1987, berry1990, lim1991}. 
In essence, superadiabatic frames emerge in an iterative procedure of progressively absorbing more of the time-dependence of $H(t)$ into the choice of basis.
The procedure is related to an asymptotic expansion and therefore has an optimal stopping point.

The ordinary adiabatic frame is the time-dependent basis of instantaneous energy eigenstate of the lab-frame Hamiltonian.
Define $U_1(t)$ to be the unitary matrix that diagonalizes the lab-frame Hamiltonian at time $t$,i.e.,
\begin{equation}
    H^{E_1}(t) = U_1^\dagger(t) H(t) U_1(t),
\end{equation}
with $H^{E_1}(t)$ diagonal. 
The superscript $E_1$ indicates the instantaneous energy basis.
The meaning of the subscripts on $U_1$ and $E_1$ will become clear momentarily.
The equation of motion in this new frame is
\begin{equation}
    i \frac{d}{dt}\rho^{E_1} = \left[ H_{\textrm{eff}}^{E_1}, \rho^{E_1} \right], \label{eq:adiabeom}
\end{equation}
where we have introduced the effective (though still Hermitian) Hamiltonian
\begin{equation}
    H_{\textrm{eff}}^{E_1} \equiv H^{E_1} + C^{E_1},
\end{equation}
and the nondiagonal (hence nonadiabatic) term
\begin{equation}
    C^{E_1}(t) \equiv - i U_1^\dagger \dot{U}_1.
\end{equation}
The adiabatic approximation assumes $C^{E_1} = 0$.

Superadiabatic frames are defined by repeating this process of instantaneous diagonalization. 
After moving into the ordinary adiabatic frame, we find the equation of motion Eq.~\eqref{eq:adiabeom}.
Although $H^{E_1}$ is diagonal, the full $H_{\textrm{eff}}^{E_1}$ is not. 
We move into the next frame by diagonalizing $H_{\textrm{eff}}^{E_1}$. 
Each iteration defines a new unitary operator $U_n(t)$ and a new effective Hamiltonian
\begin{equation}
    H_{\textrm{eff}}^{E_n} \equiv H^{E_n} + C^{{E_n}},
\end{equation}
with nonadiabatic term
\begin{equation}
    C^{E_n} \equiv -i U^\dagger_n \dot{U}_n.
\end{equation}
The subscript $n$ denotes the $n$th superadiabatic frame. 
In this numbering convention, the ordinary adiabatic frame is the 1st superadiabatic frame.

In Fig.~\ref{fig:bloch2} we re-examine the Bloch-sphere trajectory and $\Delta S (t)$ curve from the middle panel of Fig.~\ref{fig:bloch1}, now in the 1st, 2nd, and 4th superadiabatic frames instead of the lab frame.
$\Delta S$ is calculated as before except that $\bm{H}_{\textrm{eff}}^{E_n}$ is used in defining $\bm{P}^{\textrm{eq}}$ [Eq.~\eqref{eq:Peq}]. 
Entropy production is nearly monotonic in the 2nd frame.
For the choice of $\epsilon$ used in Fig.~\ref{fig:bloch2}, the 2nd frame is the optimal one in the definite quantitative sense specified below. 
In moving to the 3rd and higher frames, $\Delta S$ increasingly deviates from near-monotonic evolution.
This is particularly clear in the 4th frame, which is why it was selected instead of the 3rd frame, as the latter is more difficult to distinguish visually from the 2nd frame.

The frame-dependence of $\Delta S$ is a consequence of the fact that a time-dependent Hamiltonian does not permit the unique identification of instantaneous energy eigenstates.
The freedom to choose an effective Hamiltonian implies a freedom in defining equilibrium states. 
The ambiguity of energy eigenstates therefore has direct consequences for quantum thermodynamics. This is one of the main messages of our work.

Adiabaticity is likewise a frame-dependent concept. An adiabatic approximation can be applied in the $n$th superadiabatic frame by setting $C^{E_n} = 0$. Under this assumption, $\bm{P}$ adiabatically follows $\bm{H}_{\textrm{eff}}^{E_{n-1}}$. 
This generalizes the ordinary adiabatic approximation, in which $\bm{P}$ tracks the lab-frame Hamiltonian $\bm{H}$. 
Indeed, a time-dependent adiabaticity parameter can be defined for each frame~\cite{deschamps2008, vandermause2016}:
\begin{equation}
    Q_n(t) \equiv \frac{|\bm{H}^{E_n}(t)|}{|\bm{C}^{E_n}(t)|}.
\end{equation}
The $Q$ factor for the entire process in the $n$th superadiabatic frame is
\begin{equation}
    Q^{\textrm{min}}_n \equiv \min_{t \in (-\infty, +\infty)} Q_n(t),
\end{equation}
and the frame-independent superadiabatic $Q$ factor is
\begin{equation}
    Q \equiv \max_{n \in \lbrace 1, 2, \dots \rbrace} Q_n^{\textrm{min}}.
\end{equation}
The frame-independent adiabatic criterion, $Q \gg 1$, accounts for the possibility that a protocol may be nonadiabatic in one frame but is in fact adiabatic in another.
The optimal superadiabatic frame is the one in which $Q_n^{\textrm{min}}$ is maximal.
That is, the optimal frame is the one in which adiabaticity is most accurately assessed~\cite{deschamps2008, vandermause2016}.

\begin{figure}
    \centering
    \includegraphics[width=\linewidth]{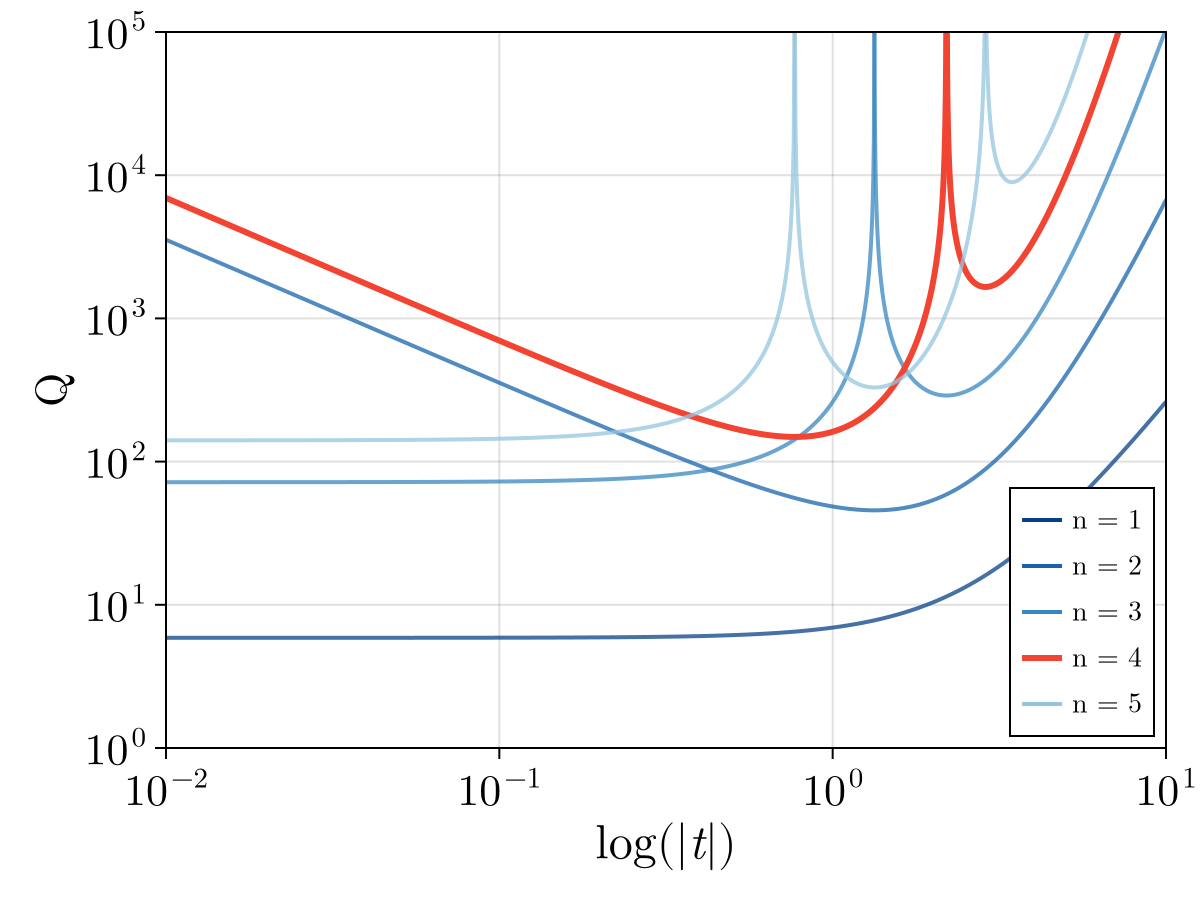}
    \caption{\textbf{\textit{Adiabaticity parameter $Q$ in different superadiabatic frames for the $\epsilon = 0.34$ Landau--Zener protocol.}}
    The $n = 4$ frame has the largest $Q^{\textrm{min}}$ and is therefore optimal.
    Only the most relevant segments of the Q evolution are displayed; $Q$ increases in all frames beyond $t=10$. 
    }
    \label{fig:AdiabaticQHistory}
\end{figure}

\begin{figure}
    \centering
    \includegraphics[width=\linewidth]{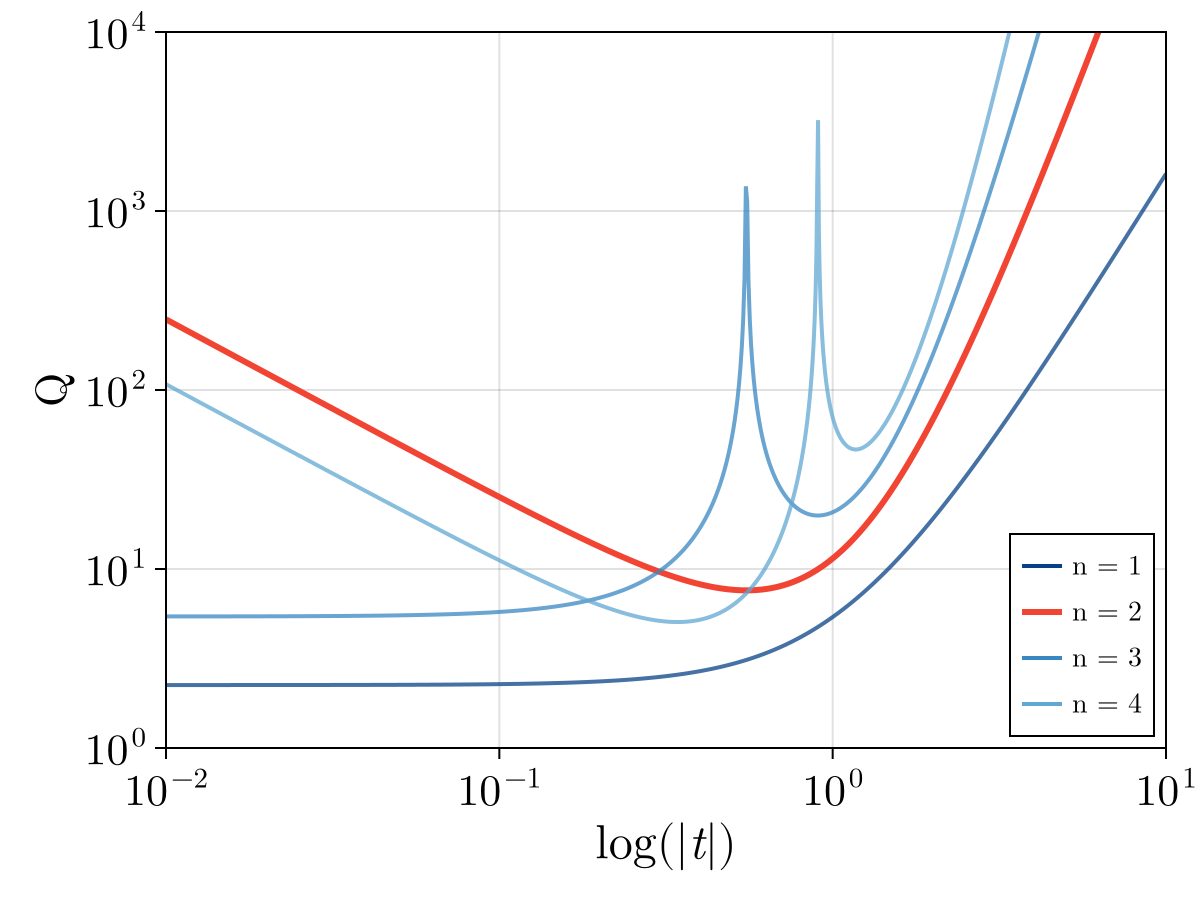}
    \caption{\textbf{\textit{Adiabaticity parameter $Q$ in different superadiabatic frames for the $\epsilon = 0.89$ Landau--Zener protocol.}}
    In this case, the optimal frame is $n = 2$, as discussed previously in connection with Fig.~\ref{fig:bloch2}. The decrease in the scale of $Q$ between Fig.~\ref{fig:AdiabaticQHistory} and the current figure reflects the decreasing adiabaticity as a function of sweep-rate parameter. 
    }
    \label{fig:DiabaticQHistory}
\end{figure}

Figure~\ref{fig:AdiabaticQHistory} shows the time-dependent adiabaticity parameter $Q_n(t)$ during a Landau--Zener sweep with $\epsilon = 0.34$ for several choices of frame index $n$.
In the 1st superadiabatic frame, $Q_1$ exhibits a single minimum at $t = 0$: nonadiabaticity principally arises from the resonance.
In moving to larger $n$, the curve is overall lifted to larger values, indicating that the adiabatic approximation is becoming more accurate. 
The time development of $Q_n$ also becomes more complicated, with additional local minima and sharp adiabatic peaks appearing.
The minimum value, $Q_n^{\textrm{min}}$, is at $t = 0$ for odd $n$, but is shifted away from $t = 0$ for even $n$.
In this case $n = 4$ is the optimal frame (red curve in Fig.~\ref{fig:AdiabaticQHistory}).

Figure~\ref{fig:DiabaticQHistory} is similar to Fig.~\ref{fig:AdiabaticQHistory} but with $\epsilon = 0.89$.
Naively this protocol appears to be away from the adiabatic limit, as suggested by $Q_1^{\textrm{min}} < 1$. 
In the $n = 2$ frame, however, we find $Q_2^{\textrm{min}} > 1$, indicating adiabatic evolution. 
In this case $n=2$ is the optimal frame.

Our findings suggest that the optimal superadiabatic frame is the one in which equilibrium and entropy are most sensibly defined, insofar as $dS/dt \geq 0$ is most nearly satisfied. 
At any given time, $Q_n$ quantifies the strength of the nonadiabatic perturbation that $\bm{P}$ is exposed to.
This is why the adiabatic approximation is best applied in the optimal frame: the perturbations that cause the $\bm{P}$ trajectory to deviate from this adiabatic path are at their smallest.
But maximally adiabatic evolution is logically distinct from maximally monotonic entropy growth.
Near-monotonicity of $\Delta S$ is an additional special attribute of the optimal superadiabatic frame.

Fundamentally, superadiabatic frames are useful for formulating quantum thermodynamics because they absorb much of the time-dependence of the lab-frame Hamiltonian into the time-dependence of the basis.
This reasoning is similar to the logic underlying the use of rotating frames in magnetic resonance and spin thermodynamics~\cite{rabi1954, abragam1958, redfield1969}.
In this respect, the use of the optimal superadiabatic frame in quantum thermodynamics generalizes the use of rotating frames to situations with more general time-dependence.

Behavior strictly consistent with the second law of thermodynamics emerges with the approximation that nonadiabaticity builds up sequentially, with $\bm{P}$ being driven away from equilibrium by $\bm{C}$ and then rethermalizing via energy dephasing.
Of course, irreversible energy dephasing is not possible under unitary dynamics, so it is unsurprising to find small non-monotonic oscillations around a path that is otherwise consistent with the second law.

\section{Arbitrary initial states\label{sec:arbitrary}}

We have seen that the coarse-grained entropy $S$ increases nearly monotonically over the course of the Landau--Zener protocol when $S$ is defined in the superadiabatic frame.
Any control protocol in which the qubit is initially in an energy eigenstate must increase the entropy from start to finish, $\Delta S (t_f) \geq 0$, because an energy eigenstate is a minimum-entropy state.
This is easy to understand geometrically.
Because $\bm{P}$ is initially (anti)aligned with $\bm{H}$, the angle between these vectors cannot decrease.
If we consider arbitrary initial states, however, $S$ may increase or decrease.

For a coherent superposition of energy eigenstates, the phase $\phi$ of the qubit (\textit{i.e.}, the azimuthal angle of $\bm{P}$ relative to $\bm{\hat{H}}$) is relevant to control. 
For example, Landau--Zener--St\"{u}ckelberg--Majorana (LZSM) interferometry refers to the interference phenomenon observed in a quantum two-level system as it is driven repeatedly through an avoided crossing~\cite{shevchenko2010, silveri2017, ivakhnenko2023}. 
It has numerous applications in areas such as quantum information processing, quantum tomography, and quantum control~\cite{oliver2005, rudner2008, berns2008, petta2010, stehlik2012, cao2013, ribeiro2013, quintana2013, gong2016, forster2014, scheuer2017, ota2018}.
In this section, we consider a single Landau--Zener sweep with an arbitrary initial state, one of the building blocks of repeated-passage driving and its applications.
In fact single-passage LZSM interferometry has itself recently been proposed as a method of coherent control, treating the initial phase as a controllable parameter used to steer a state into a desired destination~\cite{kofman2024} (see also Ref.~\cite{wubs2005}).

\begin{figure}
    \centering
    \includegraphics[width=\linewidth]{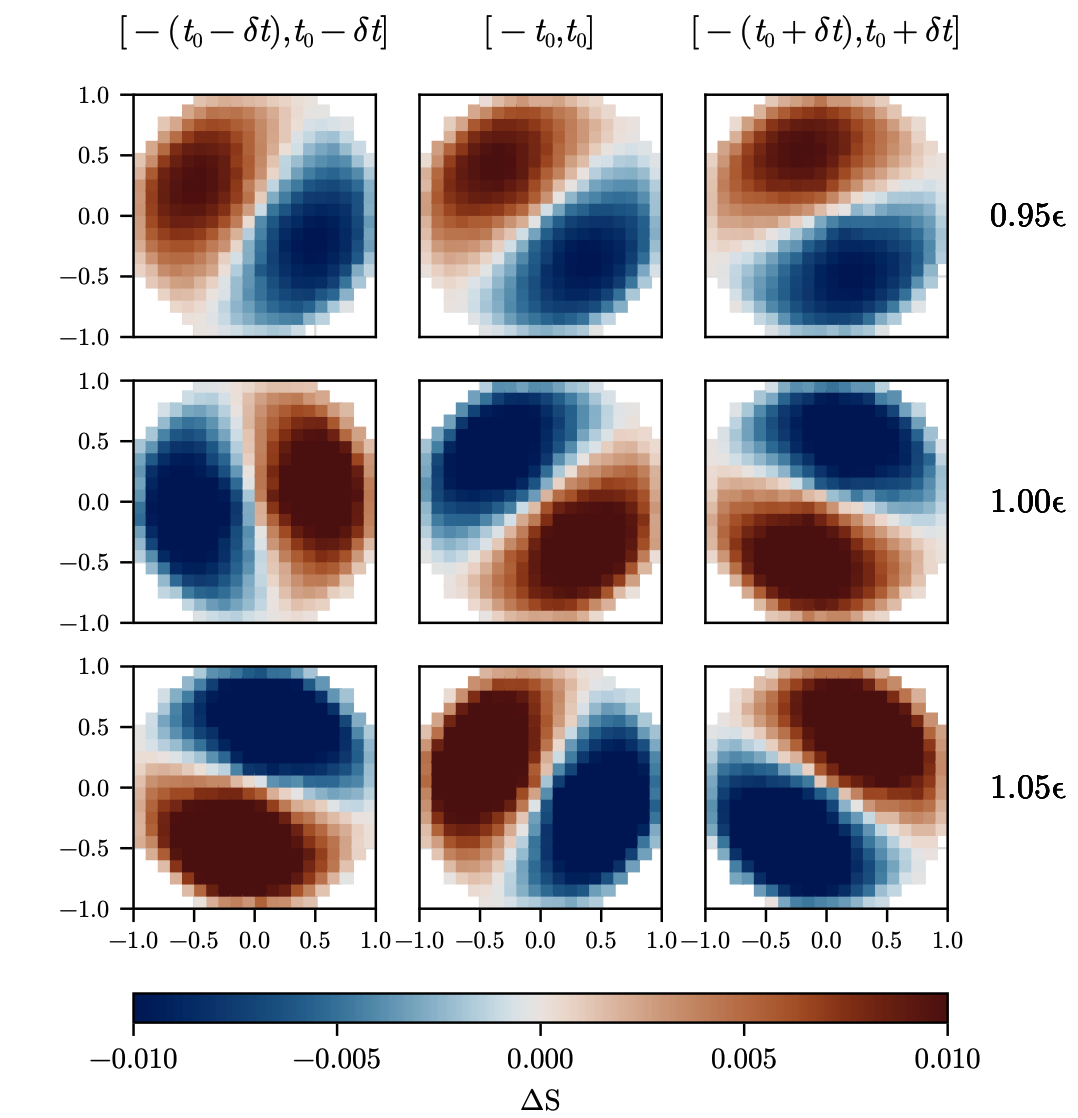}
    \caption{\textbf{\textit{Entropy production $\Delta S$ from a single Landau--Zener sweep in the adiabatic regime.}} Within each panel, the color of a cell corresponds to $\Delta S$ experienced by a qubit with an initial state in the cell.
    Numerical results are shown for reference values of $t_0 = 100 $ and $\epsilon = 0.34$ as well as for small shifts away from these values. The shift in $t_0$ is $\delta t = 0.1$. 
    }
    \label{fig:DeltaSadiab}
\end{figure}

Figure~\ref{fig:DeltaSadiab} shows $\Delta S (t_f)$ for a collection of initial states sampled over the northern hemisphere of the Bloch sphere. 
We performed numerical calculations slightly varying the sweep duration $\Delta t \equiv t_f - t_i = 2 t_f$ (the three columns) and the sweep-rate parameter $\epsilon$ (the three rows).
A dipole pattern is evident in all cases, with hot regions ($\Delta S > 0$) being separated from cold regions ($\Delta S < 0$) by a transitionless boundary ($\Delta S = 0$) running nearly through the north pole.
Qubits lying along the $\Delta S = 0$ curve experience transitionless driving, not due to counterdiabatic terms in the Hamiltonian~\cite{demirplak2003, berry2009, petiziol2024} but rather due to their initial states~\cite{kofman2024}.

\begin{figure}
    \centering
    \includegraphics[width=\linewidth]{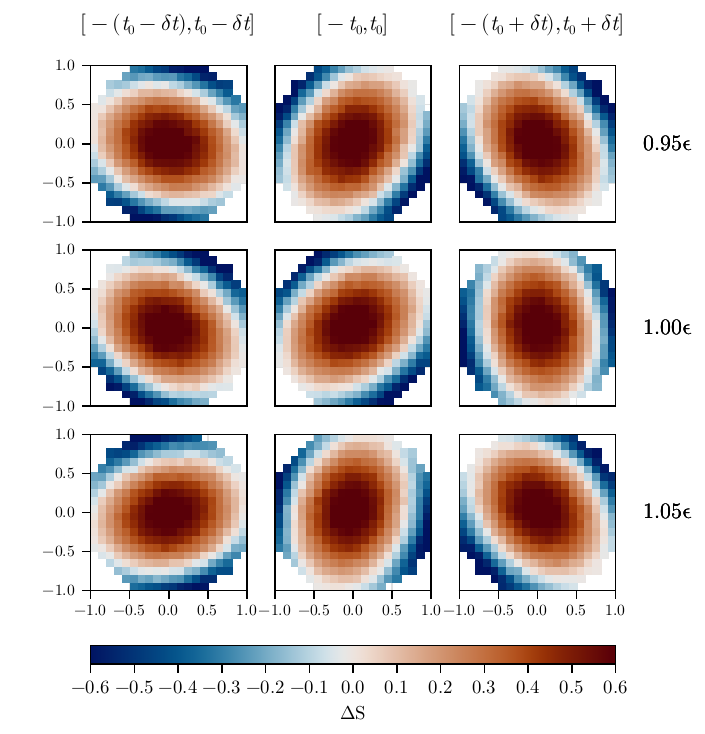}
    \caption{\textbf{\textit{Entropy production $\Delta S$ from a single Landau--Zener sweep in the diabatic regime.}} Here reference values of $t_0 = 100$ and $\epsilon = 5$ are used. The shift in $t_0$ is $\delta t = 0.1$. Compare Fig.~\ref{fig:DeltaSadiab}.
    }
    \label{fig:DeltaSdiab}
\end{figure}

The dipole pattern results from the resonance at $t = 0$, which is the dominant source of nonadiabaticity in the nearly adiabatic (small-$\epsilon$) regime.
For a given qubit, $\Delta S$ is principally determined by the orientation of $\bm{P}$ with respect to $\bm{H}$ upon entry into the resonance region. 
This interpretation is supported by the adiabatic impulse approximation, which decomposes the evolution into two adiabatic phases interrupted by a nonadiabatic impulse at the resonance~\cite{damski2005, damski2006, kofman2023}.
If $\bm{H}$ moves toward (away from) $\bm{P}$ during the impulse, then $S$ decreases (increases).

The $\Delta S$ dipole is a robust feature. Its orientation on the Bloch sphere, however, is highly sensitive to $\Delta t$ and $\epsilon$. 
This is because the position of $\bm{P}$ relative to $\bm{H}$ at resonance is dependent on all prior evolution. 
Indeed, all of the sensitivity to $\Delta t$ enters through $t_i$. We have confirmed that varying $t_f$ has no noticeable effect.
According to the adiabatic-impulse approximation, the evolution after the resonance induces no further change in $S$.

Although the $\Delta S = 0$ curve runs nearly through the north pole in Fig.~\ref{fig:DeltaSadiab}, the Landau--Zener formula tells us that it cannot pass exactly through the pole. 
For the values of $\epsilon$ used in the figure, $\Delta S_{\textrm{LZ}} \approx 0.001$. The entropy production when the initial state is an energy eigenstate is relatively small compared to other initial states.

As $\epsilon$ is increased, the $\Delta S = 0$ curve is shifted further away from the north pole.
This phenomenon can be seen in the diabatic examples of Fig.~\ref{fig:DeltaSdiab}. 
The north pole is now deep inside the $\Delta S$ hot spot. 
Sensitivity to $\Delta t$ and $\epsilon$ is still apparent, but it is less pronounced than in Fig.~\ref{fig:DeltaSadiab}.

The sudden ($\epsilon \rightarrow \infty$) limit is of limited utility in quantum control because $\bm{P}$ does not evolve appreciably during an extremely rapid sweep.
Steering is more efficacious in the adiabatic ($\epsilon \rightarrow 0$) limit, but the sensitivity to $\Delta t$ and $\epsilon$---which in practice are necessarily finite and nonzero, respectively---indicates that exquisite control is required to take advantage of the phase information of qubit states that are initially away from the poles. 

Intuitively, the relationship between the initial and final states is tenuous due to the long evolution time of a small-$\epsilon$ control protocol, assuming that $\Delta t$ is long enough to have $\bm{\hat{H}}(t_i) \approx \bm{\hat{H}}(-\infty)$ and $\bm{\hat{H}}(t_f) \approx \bm{\hat{H}}(+\infty)$. 
Nonetheless, because the dynamics is unitary, it is impossible for the initial state to be forgotten---or decorrelated from the state at a later time---in a true, fine-grained sense. Instead, the tenuousness manifests in the sensitivity of the final state to the precise control schedule.

\section{Discussion\label{sec:discussion}}

In this work, we have raised the following question: To what extent is the temporally coarse-grained entropy of a qubit a monotonically increasing function of time under coherent driving? We have analyzed the classic Landau--Zener protocol to gain insight into this question.

Our analysis encounters the general ambiguity of identifying instantaneous energy eigenstates of a time-dependent Hamiltonian. 
This point has been appreciated in other physical contexts, such as in the description of particle production in an expanding universe or a time-dependent electric field~\cite{dabrowski2014, dabrowski2016, yamada2021, alvarez2022}. 
Here, we have pointed out the significance of the ambiguity for the quantum thermodynamics of systems in which the Hamiltonian is continuing to change.

We find that entropy production $\Delta S$ is indeed nearly monotonic in the optimal superadiabatic frame. 
We leave for future work a determination of the extent to which this finding generalizes to other quantum control protocols.
Superadiabatic frames have also been employed in the derivation of master equations describing environment-induced relaxation in slowly driven systems~\cite{pekola2010, salmilehto2010, salmilehto2011}. 
The relationship between the present article and these earlier works is that our analysis focuses on irreversibility due to the slow driving itself rather than the system--environment coupling.

Unsurprisingly, slow Landau--Zener sweeps with arbitrary initial states do not generically result in $\Delta S > 0$.
However, we find that $\Delta S$ is sensitive to small fractional changes in the sweep rate and duration.
This finding is consistent with recent work on the importance of precise timekeeping in quantum control~\cite{xuereb2023}, as well as with thermodynamic intuition. 
Although entropy decreases are possible with nearly adiabatic sweeps, engineering this outcome requires fine-grained information and precise control.

\begin{acknowledgments}
PZ is supported by the David \& Lucile Packard Foundation.
LJ is supported by a Feynman Fellowship through LANL LDRD project number 20230788PRD1.
CAA are supported by the Faculty of Arts and Sciences of Harvard University, the NSF AI Institute for Artificial Intelligence and Fundamental Interactions, the Research Corporation for Science Advancement, and the David \& Lucile Packard Foundation.
\end{acknowledgments}

\bibliography{refs}
\bibliographystyle{apsrev4-2}

\end{document}